\documentclass[12pt]{article}
\usepackage{amsmath}
\usepackage{prodint}
\usepackage{graphicx}
\usepackage{enumerate}
\usepackage{natbib}
\usepackage{url} 
\usepackage{binomexp}
\usepackage{mathtools}
\usepackage{amsfonts}
\usepackage{mathrsfs}
\RequirePackage{subcaption}
\usepackage{booktabs}
\usepackage{bbm}
\usepackage{dsfont}
\usepackage{longtable}
\usepackage{tabulary}
\usepackage{yhmath}
\usepackage{bm}
\usepackage{amssymb}
\usepackage[nodisplayskipstretch]{setspace}
\usepackage{tikz}
\usetikzlibrary{arrows.meta, decorations.pathmorphing, decorations.pathreplacing, positioning, calc, shadows.blur}
\usepackage{float}
\usepackage{color}
\usepackage{stmaryrd}
\usepackage{paralist}
\usepackage{multirow}
\usepackage{threeparttable}
\usepackage{soul}
\usepackage{algorithm,algpseudocode}
\makeatletter
\newcommand{\algmargin}{\the\ALG@thistlm}
\makeatother
\newlength{\whilewidth}
\algdef{SE}[parWHILE]{parWhile}{EndparWhile}[1]
  {\parbox[t]{\dimexpr\linewidth-\algmargin}{%
     \hangindent\whilewidth\strut\algorithmicwhile\ #1\ \algorithmicdo\strut}}{\algorithmicend\ \algorithmicwhile}%
\algnewcommand{\parState}[1]{\State%
  \parbox[t]{\dimexpr\linewidth-\algmargin}{\strut #1\strut}}

\usepackage{adjustbox}
\usepackage[colorlinks=true,linkcolor=blue,citecolor=black]{hyperref}

\newtheorem{theorem}{Theorem}

\newtheorem{proposition}{Proposition}
\newtheorem{corollary}{Corollary}
\newtheorem{lemma}{Lemma}
\newtheorem{remark}{Remark}

\makeatletter
    {\@begintheorem{Example}{\ref{#1} (Continued)}}%
    {\@endtheorem}
\makeatother

\newtheorem{condition}{Condition}

\newcommand{\ind}{\mathds{1}}

\newcommand{\blind}{1}

\allowdisplaybreaks

\usetikzlibrary{shapes,decorations,arrows,calc,arrows.meta,fit,positioning}
\tikzset{
     -Latex,auto,node distance =1 cm and 1 cm,semithick,
     state/.style ={ellipse, draw, minimum width = 0.7 cm},
     point/.style = {circle, draw, inner sep=0.04cm,fill,node contents={}},
    bidirected/.style={Latex-Latex,dashed},
     el/.style = {inner sep=2pt, align=left, sloped}
 }

\begin{document}

\def\spacingset#1{\renewcommand{\baselinestretch}%
{#1}\small\normalsize} \spacingset{1}


\if1\blind
{
  \title{\Large \bf Design-based theory for covariate adjustment in generalized pairwise comparisons}
  \date{}
  \author{Xinyuan Chen$^{1,\ast}$ ~ Fan Li$^2$\vspace{0.2cm}\\
    $^1$Department of Mathematics and Statistics, \\
    Mississippi State University, Mississippi State, MS, USA \\
    $^2$Department of Biostatistics, \\
    Yale School of Public Health, New Haven, CT, USA \\
    ${}^\ast$xchen@math.msstate.edu}
  \maketitle
} \fi

\if0\blind
{
  \bigskip
  \bigskip
  \bigskip
  \begin{center}
    {\Large \bf Design-based theory for covariate adjustment in generalized pairwise comparisons}
\end{center}
  \medskip
} \fi

\begin{abstract}
Generalized pairwise comparisons summarize treatment effects in randomized experiments with ordinal, multivariate, or prioritized outcomes. We study optimal covariate adjustment under complete randomization, with potential outcomes and covariates fixed and treatment assignment as the sole source of randomness. We derive the exact randomization variance of the unadjusted estimator and characterize linear calibration adjustments using the between-arm difference in covariate means. The asymptotic variance is a strictly convex quadratic in the adjustment coefficient, yielding a closed-form optimum. The resulting estimator is asymptotically no less efficient than the unadjusted estimator, with an explicit precision gain that clarifies which covariates improve precision. However, natural pair-level and single-arm least squares adjustments generally use weights that differ from the allocation-dependent optimum and can substantially reduce efficiency. We also propose a within-arm variance estimator for asymptotically conservative inference, whose conservatism arises from unidentifiable joint variation in a unit's two potential scores, and show that residualizing each arm on its own projection coefficient saves the discrepancy that governs the efficiency loss of conventional adjustments. In the special case where the comparison function is the outcome difference, our proposed covariate-adjusted estimator is first-order equivalent to the fully interacted regression adjustment of \citet{Lin2013}.
\end{abstract}

\noindent%

\spacingset{1.75} 

\section{Introduction}

Covariate adjustment leverages pre-treatment characteristics to improve the precision of treatment effect estimation in randomized experiments. For the conventional average treatment effect estimand, design-based theory provides a rigorous justification for this practice. In particular, \citet{Lin2013} proved that the least squares estimator obtained by regressing the outcome on the treatment indicator, centered covariates, and their interactions remains consistent under the randomization distribution without requiring the working outcome model to be correctly specified; the resulting estimator is asymptotically no less efficient than the unadjusted difference in means estimator. This model-assisted efficiency guarantee has subsequently been extended to settings including cluster randomization \citep{Su2021}, rerandomization \citep{li2020rerandomization}, and missing covariates \citep{Zhao2024}. The broader lesson from this line of work is that, for estimation of the average treatment effect, linear regression adjustment is a robust design-based procedure and, when appropriately constructed, does not sacrifice asymptotic efficiency.

Endpoints in current randomized experiments are increasingly complex, often involving ordinal responses, multiple outcomes measured on different scales, or composite outcomes whose components carry different clinical priorities. In such settings, there is growing interest in obtaining a single summary of the treatment effect without reducing the endpoint to an arbitrary common scale. Generalized pairwise comparisons have emerged as a powerful framework for this purpose \citep{Buyse2010, Buyse2025}. Under this framework, the outcome of a treated unit is compared with that of a control unit according to a prespecified comparison rule, and each pair is assigned a score representing a win, loss, tie, or, more generally, the relative favorability of one outcome vector over the other. Averaging these scores across treatment-control pairs leads to estimands that include the Mann-Whitney or probabilistic index parameter, the win ratio and net benefit for prioritized composite endpoints \citep{Pocock2012, bebu2016large}, and the win odds \citep{brunner2021win}. The general formulation in terms of a comparison function on pairs of potential outcomes along with estimation through U-statistics was previously studied by \citet{Mao2018} and \citet{chen2026optimal, chen2026principal}. Covariate adjustment for these estimands has also received attention. \citet{Vermeulen2015} and \citet{Vermeulen2023} developed locally semiparametric efficient estimators of the probabilistic index; \citet{Jaspers2024} studied small-sample inference for covariate-adjusted generalized pairwise comparisons through probabilistic index models \citep{thas2012probabilistic}; \citet{Scheidegger2026} proposed standardization over fitted pairwise models for estimating the win odds; \citet{Cao2025} developed propensity score weighting and augmented estimators for the win ratio and net benefit with ordinal outcomes; and \citet{Lou2026} minimized the asymptotic variance of a calibrated Wilcoxon two-sample statistic with a continuous scalar outcome. A common feature across these developments is their reliance on a sampling-based, superpopulation framework, and most are tailored to a particular comparison function or estimand. By contrast, design-based theory conditions on the finite population of potential outcomes and covariates, treating randomized treatment assignment as the sole source of uncertainty. Within this framework, it remains unknown whether the classical validity and efficiency guarantees of linear adjustment for the average treatment effect extend to this nonlinear class of pairwise comparison estimands.

Establishing design-based theory for covariate adjustment in generalized pairwise comparisons is not straightforward for several reasons. First, generalized pairwise comparison is constructed from a collection of dependent scores. Each unit is compared repeatedly with units in the opposite group and contributes to many observed pairs that cannot be treated as independent units. Second, because the comparison is ordered, each unit can occupy two distinct positions. Specifically, it has one average score when it appears first in the comparison and another when it appears second. These position-specific average scores generally have different linear projections on the covariates, and it is not immediately clear how to combine the two projections for covariate adjustment. Third, seemingly natural regression adjustment strategies are available. For example, one may regress the pairwise score on the covariate difference, with or without treatment interactions, or regress one of the position-specific average scores on a unit's own covariates. The implied coefficients produced by these strategies are determined by the particular regression formulation and need not coincide with the optimal coefficient that minimizes randomization variance. Indeed, as we prove, these natural pair-level and unit-level adjustments can actually be less efficient than no adjustment, particularly under unequal treatment allocation. These observations raise two fundamental questions: \emph{does a linear adjustment exist that retains the design-based efficiency guarantee for general pairwise comparison estimands, and if so, how should the optimal adjustment be constructed and valid design-based inference operationalized?}

We resolve these questions by developing a unified design-based theory for covariate adjustment in generalized pairwise comparisons. We first derive the exact finite-population randomization variance of the unadjusted estimator and show that every linear adjustment operates through the between-arm difference in covariate means. Applying the exact variance formula to the residualized comparison function yields a convex quadratic leading variance whose minimizer is an allocation-weighted combination of the projections of the two position-specific average scores. Interestingly, estimating this coefficient carries no first-order cost, and the resulting feasible estimator is asymptotically no less efficient than the unadjusted estimator for any given comparison function. This result also explains why seemingly natural pair-level and unit-level regressions, whose implied weights are determined by their regression formulations rather than by the treatment allocation, may be less efficient than the unadjusted estimator. Second, we characterize the asymptotic randomization variance and propose a simple within-arm variance estimator that provides valid inference for the proposed covariate-adjusted estimator. The variance estimator is conservative due to an unidentifiable cross-world quantity. We further show that the standard heteroscedasticity-robust and one-way cluster-robust variance estimators computed from pair-level regressions are generally inconsistent. The two-way cluster-robust variance estimator consistently estimates a conservative target for a single directional comparison but is generally invalid for the net benefit unless it incorporates the covariance between the two opposite orderings; our within-arm variance construction accommodates this covariance directly and is therefore the recommended approach for operationalizing design-based inference.

\section{Generalized pairwise comparisons: A finite-population perspective}

\subsection{Set-up and estimand}\label{sec:setup}

Consider a completely randomized experiment on $N$ units in which $N_a$ units are assigned to arm $a$, with $N_0+N_1=N$ and $\pi_a=N_a/N$, and write $A_i\in\{0,1\}$ for the assignment of unit $i$ and $\ind_i(a)=\ind(A_i=a)$. Let $\bm X_i$ be a $d_x$-dimensional vector of pre-treatment covariates and $\bm Y_i=(Y_{1,i},\ldots,Y_{Q,i})^\top$ the vector of observed outcomes, with $\bm Y_i=A_i\bm Y_i(1)+(1-A_i)\bm Y_i(0)$ under the Stable Unit Treatment Value Assumption (SUTVA). We pursue a design-based framework where $\{\bm Y_i(0),\bm Y_i(1),\bm X_i\}_{i=1}^N$ are held fixed and the assignment vector, uniform on the set of vectors with $N_1$ ones, is the sole source of randomness. 

Let $w:\bm{\mathcal Y}\times\bm{\mathcal Y}\mapsto\mathbb R$ be a comparison function on the product of the outcome supports, and define $W_{i,j}(a,a')=w\{\bm Y_i(a),\bm Y_j(a')\}$, so that the observed score is $W_{i,j}=w(\bm Y_i,\bm Y_j)=\sum_{a=0}^1\sum_{a'=0}^1\ind_i(a)\ind_j(a')W_{i,j}(a,a')$. The comparison function $w$ is not necessarily symmetric, such that $W_{i,j}$ and $W_{j,i}$ are generally distinct. For later use, define the covariate difference $\bm X_{i,j}=\bm X_i-\bm X_j$, which satisfies $\bm X_{j,i}=-\bm X_{i,j}$ and $\sum_{i\neq j}\bm X_{i,j}=\bm 0$. Our primary finite-population causal estimand is defined as
\begin{equation}\label{eq:estimand}
    \lambda(a,1-a)=\frac{1}{N(N-1)}\sum_{1\leq i\neq j\leq N} w\{\bm Y_i(a),\bm Y_j(1-a)\},
\end{equation}
which is the average score over all ordered pairs of distinct units, one taken under $a$ and the other under $1-a$. The net benefit estimand is then given by $\tau(a)=\lambda(a,1-a)-\lambda(1-a,a)$. Additionally, taking $Q=1$ and $w(y_1,y_2)=y_1-y_2$ returns the conventional average treatment effect; taking $w(y_1,y_2)=\ind(y_1>y_2)+\ind(y_1=y_2)/2$ returns the probabilistic index. 

\begin{remark}\label{rmk:altestimand}
    Under the finite-population framework, one can alternatively define the causal estimand as $\lambda^{\dagger}(a,1-a)=N^{-2}\sum_{1\leq i,j\leq N}W_{i,j}(a,1-a)$ by including all unit-specific potential outcome comparisons. The difference between this estimand and \eqref{eq:estimand} can be written as
    \begin{align*}
        \lambda^{\dagger}(a,1-a)-\lambda(a,1-a)=\frac1N\left\{\frac1N\sum_{i=1}^NW_{i,i}(a,1-a)-\lambda(a,1-a)\right\},
    \end{align*}
    which is $O(N^{-1})$ whenever the unit-specific scores are bounded on average, satisfying $\limsup_NN^{-1}\sum_{i=1}^N|W_{i,i}(a,1-a)|<\infty$. Under this additional condition, all asymptotic results developed below remain unchanged if $\lambda^\dagger(a,1-a)$ is taken as the estimand. Thus, the distinction between the two estimands is asymptotically negligible. We focus on \eqref{eq:estimand} because it excludes the self-comparison $W_{i,i}(a,1-a)=w\{\bm Y_i(a),\bm Y_i(1-a)\}$, which depends jointly on two potential outcomes of the same unit and is not identified in general.
\end{remark}

\subsection{Further examples of comparison functions}\label{sec:examples}

\eqref{eq:estimand} is a general construction and covers many known estimands. As explained in \S\ref{sec:setup}, in the simplest case with a single outcome, $Q=1$, taking $w(y_1,y_2)=y_1-y_2$ makes $\lambda(a,1-a)=N^{-1}\sum_{i=1}^N\{Y_i(a)-Y_i(1-a)\}$ the average treatment effect, and taking $w(y_1,y_2)=\ind(y_1>y_2)+\ind(y_1=y_2)/2$ makes $\lambda(a,1-a)$ the finite-population probabilistic index, whose limit is ${\mathbb P}\{Y(a)>Y(1-a)\}+{\mathbb P}\{Y(a)=Y(1-a)\}/2$ and whose ratio to $\lambda(1-a,a)$ is the win odds of \citet{brunner2021win}. The same ratio formed from a contrast that does not split ties, such as the lexicographic rule below, is the win ratio of \citet{Pocock2012}.

More generally, with multiple outcomes, $Q\geq 2$, write $Y_q(a)\in\mathcal Y_q$ and $\bm{\mathcal Y}=\times_{q=1}^Q\mathcal Y_q$. When no consensus exists on the relative clinical importance of the components, we can define the following dimension-wise rule
\begin{align*}
    w(\bm y_1,\bm y_2)=\sum_{q=1}^Q\varpi_q\, w_q(y_{q,1},y_{q,2}),\quad \varpi_q\geq 0,\ \sum_{q=1}^Q\varpi_q=1,
\end{align*}
which aggregates componentwise contrasts $w_q$ such as the Heaviside score $\ind(y_1>y_2)+\ind(y_1=y_2)/2$ used above to construct a pairwise comparison estimand. When the components can be ranked by clinical importance, the lexicographic rule $w(\bm y_1,\bm y_2)=\ind(\bm y_1\succ_L\bm y_2)$, in which $\bm y_1\succ_L\bm y_2$ if the first component on which the two vectors differ favors $\bm y_1$, gives the win fraction of \citet{Pocock2012} and the net benefit of \citet{Buyse2010}. The theory below places no restriction on $w$ beyond the moment requirement of Condition \ref{cd:moments}, stated in \S\ref{sec:class}. Importantly, the comparison function $w$ need be neither symmetric, nor bounded, nor continuous. A comparison function is called antisymmetric if $w(y_1,y_2)=C-w(y_2,y_1)$ for every pair of outcome vectors and some constant $C$. The linear contrast is antisymmetric with $C=0$, and the Heaviside score with $C=1$. The strict indicator function $\ind(y_1>y_2)$ and the lexicographic rule satisfy $w(y_1,y_2)+w(y_2,y_1)=1$ for every untied pair, but both scores equal zero for an exact tie. They are therefore antisymmetric with $C=1$ when no exact ties occur. Antisymmetry is not required for Theorems \ref{thm:optimal}--\ref{thm:varest}. It is invoked only in Remark \ref{rmk:sandwich}, where it permits an explicit characterization of the failure of the two-way cluster-robust variance estimator for the net benefit, and in \S\ref{sec:netbenefit}, where it simplifies the net-benefit adjustment and variance expression.

\begin{remark}\label{rmk:nocensor}
    Our theory requires the comparison function $w$ to be evaluable for every ordered pair of potential outcomes and therefore does not directly cover right-censored prioritized endpoints, for which a comparison may remain unresolved because follow-up ends before the pair can be classified; see, for example, \citet{fang2026making}. In that setting, $\lambda(a,1-a)$ is not identified without additional assumptions. Although inverse probability of censoring weighting can address this issue \citep{Dong2020}, it reweights the pairwise contributions and therefore falls outside the class of estimators characterized in Lemma \ref{lem:class}.
\end{remark}

\subsection{The unadjusted estimator and its exact randomization variance}\label{sec:unadjusted}

Let $\mathcal S(a,1-a)=\{(i,j): A_i=a,\ A_j=1-a,\ 1\leq i\neq j\leq N\}$, a set of $N_aN_{1-a}$ ordered pairs, and let
\begin{equation}\label{eq:unadjusted}
    \widehat\lambda(a,1-a)=\frac{1}{N_aN_{1-a}}\sum_{(i,j)\in\mathcal S(a,1-a)}W_{i,j}
\end{equation}
be the unadjusted estimator, obtained equivalently as the coefficient of $\ind_i(a)\ind_j(1-a)$ in the least squares fit of $W_{i,j}$ on the two indicators $\{\ind_i(a)\ind_j(1-a),\ \ind_i(1-a)\ind_j(a)\}$ without an intercept, over the $N(N-1)$ ordered pairs. Of note, this is the pairwise analog of the difference in means. We then write $\epsilon_{i,j}(a,1-a)=W_{i,j}(a,1-a)-\lambda(a,1-a)$ for the centered potential score and define the two \emph{position-specific scores} of unit $i$ as
\begin{equation}\label{eq:rolescores}
    \overline\epsilon_{i,\cdot}(a,1-a)=\frac{1}{N-1}\sum_{j:j\neq i}\epsilon_{i,j}(a,1-a),\qquad
    \overline\epsilon_{\cdot,i}(a,1-a)=\frac{1}{N-1}\sum_{j:j\neq i}\epsilon_{j,i}(a,1-a).
\end{equation}
The first is the centered average score of unit $i$ when it is placed in arm $a$ and compared with the remaining units placed in arm $1-a$, and the second is the centered average score accruing to those units when $i$ is instead placed in arm $1-a$. When the comparison function is the strict indicator function, i.e., $w(y_1,y_2)=\ind(y_1>y_2)$, the first is the centered fraction of comparisons that $i$ wins and the second the centered fraction that it loses. Both are fixed finite-population quantities satisfying $\sum_{i=1}^N\overline\epsilon_{i,\cdot}(a,1-a)=\sum_{i=1}^N\overline\epsilon_{\cdot,i}(a,1-a)=0$. Then, for a pair of sequences $\mathcal E_i=(\mathcal E_{i,\cdot},\mathcal E_{\cdot,i})$ that each sum to zero over $i$, we define the following variance operators
\begin{equation}\label{eq:Vdef}
    V_c(\mathcal E)=\frac{1}{\pi_aN}\sum_{i=1}^N\mathcal E_{i,\cdot}^2+\frac{1}{\pi_{1-a}N}\sum_{i=1}^N\mathcal E_{\cdot,i}^2,\qquad
    V(\mathcal E)=V_c(\mathcal E)-\frac{1}{N}\sum_{i=1}^N\left(\mathcal E_{i,\cdot}+\mathcal E_{\cdot,i}\right)^2 .
\end{equation}
The randomization variance of \eqref{eq:unadjusted} is available in closed form for every finite population size $N$. We define the following intermediate quantities before stating this result:
\begin{equation}\label{eq:RQP}
\begin{aligned}
    &Q_\epsilon=\sum_{i\neq j}\epsilon_{i,j}(a,1-a)^2,\qquad P_\epsilon=\sum_{i\neq j}\epsilon_{i,j}(a,1-a)\epsilon_{j,i}(a,1-a),\\
    &R_1=(N-1)^2\sum_{i=1}^N\overline\epsilon_{i,\cdot}(a,1-a)^2,\quad
    R_2=(N-1)^2\sum_{i=1}^N\overline\epsilon_{\cdot,i}(a,1-a)^2,\\
    &R_3=(N-1)^2\sum_{i=1}^N\overline\epsilon_{i,\cdot}(a,1-a)\overline\epsilon_{\cdot,i}(a,1-a).
\end{aligned}
\end{equation}

\begin{proposition}\label{prop:exact}
    Under complete randomization with $N\geq 4$, the estimator \eqref{eq:unadjusted} is unbiased, $\mathbb E\{\widehat\lambda(a,1-a)\}=\lambda(a,1-a)$, and its exact randomization variance is
    \begin{align*}
        {\rm var}\{\widehat\lambda(a,1-a)\}=\frac{\Pi}{N_a^2N_{1-a}^2}\Big\{&(N_{1-a}-1)(N_{1-a}-2)R_1+(N_a-1)(N_a-2)R_2 \displaybreak[0]\\
        &-2(N_a-1)(N_{1-a}-1)R_3+(N_a-1)(N_{1-a}-1)(Q_\epsilon+P_\epsilon)\Big\},
    \end{align*}
    where $\Pi=N_aN_{1-a}\{N(N-1)(N-2)(N-3)\}^{-1}$.
\end{proposition}

Proposition \ref{prop:exact} is an algebraic identity, proved in \S{S3} of the Supplementary Material by enumerating the seven possible overlap patterns of two ordered pairs and evaluating the corresponding hypergeometric moments of the assignment indicators. Here, the terms in $R_1$ and $R_2$ carry the two position-specific scores with weights determined by the sample size in each arm, and dominate the variance; the term in $R_3$ is a covariance between the two scores of the same unit. The term in $Q_\epsilon+P_\epsilon$ of \eqref{eq:RQP} collects the pairs that share both indices, and is of smaller order whenever the position-specific scores do not all vanish, because $Q_\epsilon$ and $P_\epsilon$ are sums of $N(N-1)$ terms whereas $R_1$, $R_2$ and $R_3$ are sums of $N$ terms each multiplied by $(N-1)^2$. The exception is the degenerate case with $R_1=R_2=R_3=0$, or, equivalently, when both position-specific average scores vanish for every unit. In this case, the order-$N^{-1}$ component disappears, and the variance is instead determined by $Q_\epsilon+P_\epsilon$ at order $N^{-2}$. The estimator consequently has a different convergence rate, and its limiting distribution is generally a quadratic form in the assignment indicators rather than the Gaussian limit developed below. We exclude this degenerate case from the subsequent asymptotic analysis. Finally, the exact formula also identifies the leading term of the randomization variance. When the treatment proportions remain bounded away from zero and one and the finite-population second moments of the centered pairwise scores remain bounded (Conditions \ref{cd:design} and \ref{cd:moments} stated in \S\ref{sec:class}), Proposition \ref{prop:exact} yields
\begin{equation}\label{eq:leading}
    N\,{\rm var}\{\widehat\lambda(a,1-a)\}=V\{\overline\epsilon_i(a,1-a)\}+O(N^{-1}),
\end{equation}
where $\overline\epsilon_i(a,1-a)=\{\overline\epsilon_{i,\cdot}(a,1-a),\overline\epsilon_{\cdot,i}(a,1-a)\}$ and $V$ is the scalar variance functional defined in \eqref{eq:Vdef}. Thus, the two position-specific average scores determine the randomization variance up to an error of order $N^{-2}$. This conclusion follows directly from the exact variance formula, without requiring an asymptotic decomposition of the estimator; the extraction is carried out in \S{S4}.

\section{Covariate adjustment via optimal calibration}

\subsection{The class of linear adjustments}\label{sec:class}

The following regularity conditions are used for the distributional results. To state these regularity conditions, we first write $\overline{\bm X}_{i,\cdot}=(N-1)^{-1}\sum_{j\neq i}\bm X_{i,j}=\{N/(N-1)\}(\bm X_i-\overline{\bm X})$ and $\overline{\bm X}_{\cdot,i}=-\overline{\bm X}_{i,\cdot}$, and $\bm S_{XX}=N^{-1}\sum_{i=1}^N\overline{\bm X}_{i,\cdot}\overline{\bm X}_{i,\cdot}^\top=\{N/(N-1)\}^2N^{-1}\sum_{i=1}^N(\bm X_i-\overline{\bm X})(\bm X_i-\overline{\bm X})^\top$. 

\begin{condition}\label{cd:design}
    The proportion $\pi_a$ has a limit in $(0,1)$ for $a=0,1$, and $d_x$ is fixed and does not depend on $N$.
\end{condition}

\begin{condition}\label{cd:moments}
    For $a',a''=0,1$, $\{N(N-1)\}^{-1}\sum_{i\neq j}W_{i,j}(a',a'')^4=O(1)$, and $N^{-1}\sum_{i=1}^N\|\overline{\bm X}_{i,\cdot}\|^4=O(1)$.
\end{condition}

\begin{condition}\label{cd:nondeg}
    The limit of $\bm S_{XX}$ exists and is positive definite, and
    \begin{align*}
        \max_{1\leq i\leq N}\left\{\overline\epsilon_{i,\cdot}(a,1-a)^2+\overline\epsilon_{\cdot,i}(a,1-a)^2+\|\overline{\bm X}_{i,\cdot}\|^2\right\}\Big/ N\longrightarrow 0 .
    \end{align*}
\end{condition}

Conditions \ref{cd:design}-\ref{cd:nondeg} play the same role as the regularity conditions of \citet{Lin2013} and \citet{Li2017} in finite-population contexts, but restated for position-specific scores rather than observed outcomes. Condition \ref{cd:nondeg} is the Lindeberg-type form of \citet{Li2017} rather than the moment-convergence form of \citet{Lin2013}. Condition \ref{cd:moments} bounds average fourth moments and rules out a small number of pairs dominating the sum. Condition \ref{cd:nondeg} supplies the Lindeberg-type requirement of the finite-population central limit theorem; its displayed part is in fact implied by Condition \ref{cd:moments}, since a bounded average fourth moment bounds the largest squared term by $O(N^{1/2})$, and it is stated to name the hypothesis the argument uses. Two further hypotheses are carried by the distributional results and stated where they are used. The first is that the limiting variance be bounded away from zero, which excludes the degenerate case described in \S\ref{sec:unadjusted}. The second is that the variance sequence converges and is not just bounded, which the finite-population central limit theorem of \citet{Li2017} also requires and which Conditions \ref{cd:design}--\ref{cd:nondeg} alone do not guarantee; without it the limiting distribution may differ across subsequences, although the confidence interval of Theorem \ref{thm:varest} remains asymptotically conservative, by the subsequence argument given in \S{S7}.

Let $\overline{\bm X}_{(a)}=N_a^{-1}\sum_{i=1}^N\ind_i(a)\bm X_i$ and let
\begin{align*}
    \overline{\bm X}_{\mathcal S(a,1-a)}=\frac{1}{N_aN_{1-a}}\sum_{(i,j)\in\mathcal S(a,1-a)}\bm X_{i,j}=\overline{\bm X}_{(a)}-\overline{\bm X}_{(1-a)}
\end{align*}
be the observed between-arm covariate imbalance. For $\bm c\in\mathbb R^{d_x}$ define the calibrated estimator
\begin{equation}\label{eq:calibrated}
    \widehat\lambda(a,1-a;\bm c)=\widehat\lambda(a,1-a)-\overline{\bm X}_{\mathcal S(a,1-a)}^\top\bm c .
\end{equation}

\begin{lemma}\label{lem:class}
    The estimator \eqref{eq:calibrated} is the unadjusted estimator \eqref{eq:unadjusted} computed from the residualized comparison function $w_{\bm c}\{(\bm y_1,\bm x_1),(\bm y_2,\bm x_2)\}=w(\bm y_1,\bm y_2)-\bm c^\top(\bm x_1-\bm x_2)$, and the estimand is unchanged, so $\mathbb E\{\widehat\lambda(a,1-a;\bm c)\}=\lambda(a,1-a)$, for every nonrandom $\bm c$. Moreover, any estimator of the form
    $\widehat\lambda(a,1-a)-(\overline{\bm X}_{(a)}-\overline{\bm X})^\top\bm c_1+(\overline{\bm X}_{(1-a)}-\overline{\bm X})^\top\bm c_0$
    with arm-specific coefficients $\bm c_1$ and $\bm c_0$ is identical to \eqref{eq:calibrated} with $\bm c=\pi_{1-a}\bm c_1+\pi_a\bm c_0$.
\end{lemma}

Lemma \ref{lem:class} is proved in \S{S6}. Two immediate observations follow from it. First, the exact unbiasedness holds for every nonrandom $\bm c$, because $\sum_{i\neq j}\bm X_{i,j}=\bm 0$ makes $\mathbb E(\overline{\bm X}_{\mathcal S(a,1-a)})=\bm 0$. It is a property of a fixed coefficient and not of the calibration itself; \S\ref{sec:feasible} addresses the consequence when the coefficient is estimated. Second, \eqref{eq:calibrated} is an estimator of the form \eqref{eq:unadjusted}, but now with $W_{i,j}(a,1-a)$ replaced by $W_{i,j}(a,1-a)-\bm c^\top\bm X_{i,j}$, and the position-specific scores \eqref{eq:rolescores} replaced by the calibration residuals
\begin{equation}\label{eq:residuals}
    \overline r_{i,\cdot}(a,1-a;\bm c)=\overline\epsilon_{i,\cdot}(a,1-a)-\overline{\bm X}_{i,\cdot}^\top\bm c,\qquad
    \overline r_{\cdot,i}(a,1-a;\bm c)=\overline\epsilon_{\cdot,i}(a,1-a)+\overline{\bm X}_{i,\cdot}^\top\bm c.
\end{equation}

\subsection{The optimal coefficient and efficiency guarantee}\label{sec:optimal}

Define the two finite-population least squares coefficients
\begin{equation}\label{eq:gammas}
    \bm\gamma_1=\bm S_{XX}^{-1}\frac{1}{N}\sum_{i=1}^N\overline{\bm X}_{i,\cdot}\,\overline\epsilon_{i,\cdot}(a,1-a),\qquad
    \bm\gamma_2=\bm S_{XX}^{-1}\frac{1}{N}\sum_{i=1}^N\overline{\bm X}_{\cdot,i}\,\overline\epsilon_{\cdot,i}(a,1-a),
\end{equation}
where the argument $(a,1-a)$ is suppressed on the left, obtained by regressing each of the two position-specific scores of \eqref{eq:rolescores} on the corresponding centered covariate vector across the whole finite population. We obtain the following variance result for the calibrated residuals.

\begin{theorem}\label{thm:optimal}
    Let $\bm S_{XX}$ be nonsingular. Then, for every $\bm c$,
    \begin{equation}\label{eq:quadratic}
        V\{\overline r_i(a,1-a;\bm c)\}=V\{\overline\epsilon_i(a,1-a)\}-\frac{1}{\pi_a\pi_{1-a}}\left\{\bm c^{\ast\top}\bm S_{XX}\bm c^{\ast}-(\bm c-\bm c^{\ast})^\top\bm S_{XX}(\bm c-\bm c^{\ast})\right\},
    \end{equation}
    where
    \begin{equation}\label{eq:cstar}
    \bm c^{\ast}=\pi_{1-a}\bm\gamma_1(a,1-a)+\pi_{a}\bm\gamma_2(a,1-a).
    \end{equation}
    Consequently, $\bm c^{\ast}$ of \eqref{eq:cstar} is the unique minimizer of $V\{\overline r_i(a,1-a;\bm c)\}$, and
    \begin{align*}
        V\{\overline r_i(a,1-a;\bm c^{\ast})\}=V\{\overline\epsilon_i(a,1-a)\}-\frac{1}{\pi_a\pi_{1-a}}\bm c^{\ast\top}\bm S_{XX}\bm c^{\ast}\leq V\{\overline\epsilon_i(a,1-a)\},
    \end{align*}
    with equality if and only if $\bm c^{\ast}=\bm 0$. If in addition Conditions \ref{cd:design}--\ref{cd:nondeg} hold, $\liminf_{N\rightarrow\infty}\allowbreak V\{\overline r_i(a,1-a;\bm c)\}>0$, and $V\{\overline r_i(a,1-a;\bm c)\}$ together with the two within-arm second moments composing it converge, then $N^{1/2}\{\widehat\lambda(a,1-a;\bm c)-\lambda(a,1-a)\}$ converges weakly to the normal distribution with mean zero and variance equal to $\lim_{N\rightarrow\infty} V\{\overline r_i(a,1-a;\bm c)\}$.
\end{theorem}

Theorem \ref{thm:optimal} is proved in \S{S7}; the main result is the exact finite-population identity \eqref{eq:quadratic}, and asymptotic normality rests on the linear approximation in the assignment indicators of \S{S5}, whose error has variance $O(N^{-2})$ by Proposition \ref{prop:exact}, so that the finite-population central limit theorem of \citet{Li2017} applies to the linear term.

\begin{remark}\label{rmk:exact}
    Theorem \ref{thm:optimal} minimizes the leading term $V$. The exact randomization variance of Proposition \ref{prop:exact}, applied to the residualized contrast $w_{\bm c}$, is itself a strictly convex quadratic in $\bm c$ with the closed-form minimizer $\bm c^{\dagger}=\{(N_{1-a}-1)\bm\gamma_1+(N_a-1)\bm\gamma_2\}/(N-2)$ for every $N\geq 4$, and $\bm c^{\dagger}-\bm c^{\ast}=(\pi_{1-a}-\pi_a)(\bm\gamma_1-\bm\gamma_2)/(N-2)$, so the two have the same first-order asymptotic properties and we use $\bm c^{\ast}$ throughout; the derivation is in \S{S8}.
\end{remark}

Several further implications from Theorem \ref{thm:optimal} are worth mentioning. First, the optimally calibrated estimator is never asymptotically less efficient than the unadjusted estimator, regardless of the comparison function, outcome dimension, or allocation ratio. Second, the precision gain is $(\pi_a\pi_{1-a})^{-1}$ times the variance explained by regressing the allocation-weighted combination $\pi_{1-a}\overline\epsilon_{i,\cdot}(a,1-a)-\pi_a\overline\epsilon_{\cdot,i}(a,1-a)$ of a unit's two position-specific scores on the covariates. In this setting, the prognostic value of covariates for precision is determined by their ability to predict the allocation-weighted combination of the two position-specific average scores. Strong prediction of each score separately may not improve efficiency. In fact, the gain vanishes when $\pi_{1-a}\bm\gamma_1=-\pi_a\bm\gamma_2$, because the two contributions cancel. Conversely, even modest predictive associations with the individual scores can improve efficiency when their allocation-weighted combination is nonzero. Third, the map $\bm c\mapsto V\{\overline r_i(a,1-a;\bm c)\}$ is a quadratic with a single minimum, and a given adjustment beats no adjustment if and only if it lies strictly inside the ellipsoid $(\bm c-\bm c^\ast)^\top\bm S_{XX}(\bm c-\bm c^\ast)<\bm c^{\ast\top}\bm S_{XX}\bm c^\ast$ centered at $\bm c^\ast$ and passing through the origin. Hence, for the generalized comparison estimand, covariate adjustment is not automatically beneficial; only covariate adjustment with the right implied coefficient becomes beneficial. Figure \ref{fig:geometry} gives a geometric illustration.

\begin{corollary}\label{cor:lin}
    Let $Q=1$ and $w(y_1,y_2)=y_1-y_2$, so that $\lambda(1,0)$ is the conventional average treatment effect. Then $\overline\epsilon_{i,\cdot}(1,0)=Y_i(1)-\overline Y(1)+O(N^{-1})$ and $\overline\epsilon_{\cdot,i}(1,0)=-\{Y_i(0)-\overline Y(0)\}+O(N^{-1})$, and therefore $\bm\gamma_1(1,0)$ and $\bm\gamma_2(1,0)$ are the finite-population least squares coefficients of $Y_i(1)$ and of $Y_i(0)$ on $\bm X_i$ up to $O(N^{-1})$. Consequently, $\bm c^\ast=\pi_0\bm\beta_1+\pi_1\bm\beta_0+O(N^{-1})$, where $\bm\beta_a$ is the finite-population least squares coefficient of $Y_i(a)$ on $\bm X_i$. Under the stated regularity conditions, $\widehat\lambda(1,0;\bm c^\ast)$ differs from the fully interacted least squares estimator of \citet{Lin2013} by $O_{\mathbb P}(N^{-1})$. Thus, the optimal adjustment recovers Lin's estimator to first order.
\end{corollary}

Corollary \ref{cor:lin} is proved in \S{S9}. It shows that our results extend the design-based theory of \citet{Lin2013} from the average treatment effect to a broader class of causal estimands defined through general pairwise comparison functions. The central efficiency guarantee continues to hold. That is, an appropriately constructed linear adjustment is asymptotically no less efficient than the unadjusted estimator. The construction of that adjustment, however, depends on the comparison function. For the simplest difference contrast, the optimal coefficient reduces to the familiar allocation-weighted combination of the arm-specific outcome regression coefficients, which is consistently estimated by fully interacted least squares. For general pairwise comparisons, the optimal coefficient instead combines the covariate projections of the two position-specific average scores rather than those of the individual outcomes.

\subsection{Why seemingly natural regression adjustment is not guaranteed optimal}\label{sec:existing}

We now examine two seemingly natural regression adjustment strategies for generalized pairwise comparisons, one at the pair level and the other at the unit level, that mimic conventional least squares adjustment for the average treatment effect. We show that these constructions need not recover the optimal adjustment coefficient and are not generally guaranteed to improve or preserve asymptotic efficiency. At the pair level, one regresses the observed score on the covariate difference, either as a main effect,
\begin{equation}\label{eq:pairancova}
    W_{i,j}\sim \left\{\ind_i(a)\ind_j(1-a),\ \ind_i(1-a)\ind_j(a),\ \bm X_{i,j}\right\},
\end{equation}
or with treatment interactions,
\begin{equation}\label{eq:pairlin}
    W_{i,j}\sim\left\{\ind_i(a)\ind_j(1-a),\ \ind_i(1-a)\ind_j(a),\ \ind_i(a)\ind_j(1-a)\bm X_{i,j},\ \ind_i(1-a)\ind_j(a)\bm X_{i,j}\right\}.
\end{equation}
Both are fitted without an intercept over the $N(N-1)$ ordered pairs, and the resulting estimator is the coefficient of $\ind_i(a)\ind_j(1-a)$. Specification \eqref{eq:pairlin} is the pairwise transcription of the fully interacted regression of \citet{Lin2013}, and \eqref{eq:pairancova} is its analysis of covariance counterpart. These specifications are related to the pairwise regression models of \citet{DeNeve2015} and \citet{thas2012probabilistic}, but use an identity link to target a marginal estimand. At the unit level, one forms, for each unit, its observed position-specific score against the opposite arm and regresses it on that unit's own covariates within its arm. Writing $\widehat W_{i,\cdot}$ and $\widehat W_{\cdot,i}$ for the two observed position-specific scores and $\widehat{\bm\gamma}_1$ and $\widehat{\bm\gamma}_2$ for the resulting arm-specific coefficients, both defined in \S\ref{sec:feasible}, where the second carries the sign of the regressor $\overline{\bm X}_{\cdot,i}$, the two unit-level estimators are $\widehat\lambda(a,1-a)-\overline{\bm X}_{\mathcal S(a,1-a)}^\top\widehat{\bm\gamma}_1$ and $\widehat\lambda(a,1-a)-\overline{\bm X}_{\mathcal S(a,1-a)}^\top\widehat{\bm\gamma}_2$. This formulation simplifies regression fit for large $N$ by using unit-level summaries instead of all $N(N-1)$ ordered pairs.

\begin{proposition}\label{prop:family}
    Under Conditions \ref{cd:design}--\ref{cd:nondeg}, each of these estimators is of the form \eqref{eq:calibrated}, and for any $\bm c$,
    \begin{equation}\label{eq:gap}
        V\{\overline r_i(a,1-a;\bm c)\}-V\{\overline r_i(a,1-a;\bm c^{\ast})\}=\frac{1}{\pi_a\pi_{1-a}}(\bm c-\bm c^{\ast})^\top\bm S_{XX}(\bm c-\bm c^{\ast}).
    \end{equation}
    The coefficient of the fully interacted pairwise regression \eqref{eq:pairlin} differs by $o_{\mathbb P}(1)$ from its finite-population counterpart, which equals $\tfrac12(\bm\gamma_1+\bm\gamma_2)$ for every $N$. The coefficients of the two unit-level regressions differ from $\bm\gamma_1$ and from $\bm\gamma_2$ by $o_{\mathbb P}(1)$. All three therefore lie on the segment
    \begin{align*}
        \bm c(\omega)=(1-\omega)\bm\gamma_1+\omega\,\bm\gamma_2,\qquad \omega\in[0,1],
    \end{align*}
    at $\omega=1/2$, $\omega=0$ and $\omega=1$ respectively. With a slight abuse of notation, $\bm c(\omega)$ denotes both this coefficient and the corresponding member of the family of estimators $\widehat\lambda\{a,1-a;\bm c(\omega)\}$ of \eqref{eq:calibrated}, so that a statement about the efficiency of $\bm c(\omega)$ is a statement about the estimator it defines. Writing $\|\bm v\|^2_{\bm S_{XX}}=\bm v^\top\bm S_{XX}\bm v$, the excess randomization variance over the optimum is
    \begin{equation}\label{eq:excess}
        V\{\overline r_i(a,1-a;\bm c(\omega))\}-V\{\overline r_i(a,1-a;\bm c^{\ast})\}=\frac{(\omega-\pi_a)^2}{\pi_a\pi_{1-a}}\left\|\bm\gamma_1-\bm\gamma_2\right\|^2_{\bm S_{XX}},
    \end{equation}
    and $\bm c(\omega)$ is asymptotically less efficient than no adjustment at all if and only if
    \begin{equation}\label{eq:hurts}
        |\omega-\pi_a|\cdot\left\|\bm\gamma_1-\bm\gamma_2\right\|_{\bm S_{XX}}>\left\|\pi_{1-a}\bm\gamma_1+\pi_a\bm\gamma_2\right\|_{\bm S_{XX}},
    \end{equation}
    in which the left-hand side is $|\omega-\pi_a|$ times the length of $\bm\gamma_1-\bm\gamma_2$. The coefficient of the pairwise analysis of covariance \eqref{eq:pairancova} converges instead to
    \begin{align*}
        \pi_a\pi_{1-a}\left\{\bm\delta(a,1-a)+\bm\delta(1-a,a)\right\}+\pi_a^2\bm\delta(a,a)+\pi_{1-a}^2\bm\delta(1-a,1-a),
    \end{align*}
    where $\bm\delta(a',a'')$ is the finite-population least squares coefficient of $W_{i,j}(a',a'')$ on $\bm X_{i,j}$, so that it depends on the within-arm potential contrasts $W_{i,j}(a',a')$ and in general lies off the segment.
\end{proposition}

\begin{figure}
    \centering
    \includegraphics{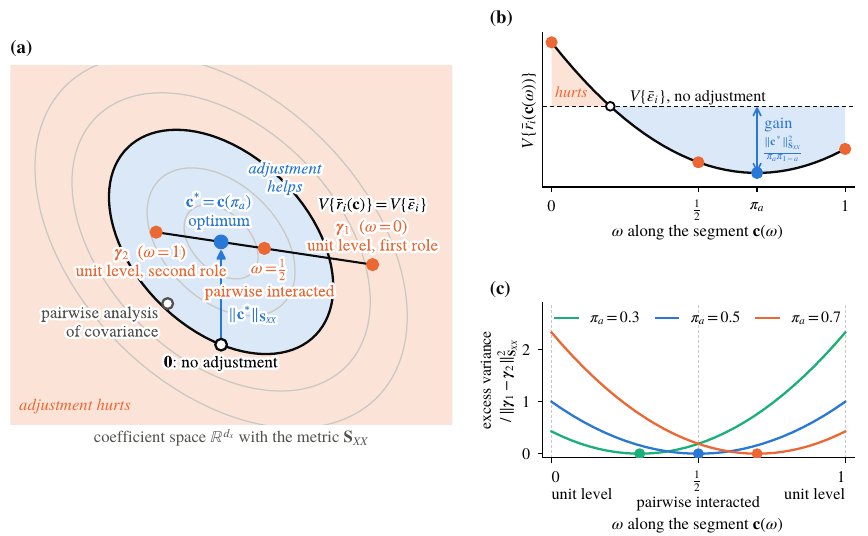}
    \caption{The geometry of \S\S\ref{sec:class}--\ref{sec:existing} in the coefficient space, drawn for $d_x=2$ and $\pi_a=0.7$ with illustrative values of $\bm\gamma_1$, $\bm\gamma_2$ and $\bm S_{XX}$. (a) The variance $V\{\overline r_i(a,1-a;\bm c)\}$ in Theorem \ref{thm:optimal} is a quadratic in $\bm c$ whose level sets are the grey ellipses centered at $\bm c^{\ast}$. The bold ellipse passes through the origin, which is no adjustment, so a coefficient is more efficient than no adjustment when it lies inside. The two unit-level regressions, at $\bm\gamma_1$ and $\bm\gamma_2$, and the fully interacted pairwise regression, at their midpoint, lie on the segment $\bm c(\omega)$ in Proposition \ref{prop:family} at the fixed positions $\omega=0$, $1$ and $1/2$. The optimum lies on the same segment at $\omega=\pi_a$. Here $\bm\gamma_1$ lies outside the bold ellipse, so the first unit-level regression is less efficient than no adjustment. The pairwise analysis of covariance lies off the segment. (b) The same variance along the segment is a parabola in $\omega$ with vertex at $\pi_a$. The dashed line is the variance without adjustment, and the shaded region where the curve lies above it is where \eqref{eq:hurts} holds. (c) The excess variance \eqref{eq:excess} in units of $\|\bm\gamma_1-\bm\gamma_2\|^2_{\bm S_{XX}}$ at three allocations. The vertex moves with $\pi_a$, and the three regressions do not, so the fully interacted pairwise regression is optimal only at $\pi_a=1/2$, and the unit-level regressions are never optimal unless $\bm\gamma_1=\bm\gamma_2$.}
    \label{fig:geometry}
\end{figure}

Proposition \ref{prop:family} is proved in \S{S10}. It shows that fully interacted pair-level regression is optimal when treatment allocation is balanced. Otherwise, by \eqref{eq:excess} its excess variance over the optimum is $\{(\pi_a-1/2)^2/(\pi_a\pi_{1-a})\}\|\bm\gamma_1-\bm\gamma_2\|^2_{\bm S_{XX}}$, and by \eqref{eq:hurts} the estimator can be less efficient than no adjustment at all once the allocation is sufficiently unbalanced relative to the gain available. The unit-level regressions use an endpoint of the segment rather than an interior point (Figure \ref{fig:geometry}). Since $\pi_a$ lies strictly inside $(0,1)$, they are suboptimal at every allocation unless $\bm\gamma_1=\bm\gamma_2$, in which case the segment degenerates to a point and every member of the family is optimal. By \eqref{eq:hurts}, the first unit-level regression is less efficient than no adjustment whenever $\pi_a\|\bm\gamma_1-\bm\gamma_2\|_{\bm S_{XX}}$ exceeds $\|\pi_{1-a}\bm\gamma_1+\pi_a\bm\gamma_2\|_{\bm S_{XX}}$, and the second whenever the same holds with $\pi_a$ replaced by $\pi_{1-a}$. Under equal allocation, both criteria reduce to $\bm\gamma_1^\top\bm S_{XX}\bm\gamma_2<0$. By the sign convention in \eqref{eq:gammas}, this corresponds to positive finite-population covariance between the covariate projections of the two position-specific scores. Thus, unit-level adjustment reduces efficiency when these projections are positively associated and improves efficiency when they are negatively associated.

The pairwise analysis of covariance in \eqref{eq:pairancova} introduces an additional complication. Because the covariate difference enters as a main effect, pairs from the same treatment arm also contribute to the fitted adjustment coefficient. Its probability limit thus depends on the within-arm potential contrasts $W_{i,j}(a',a')$, which do not enter the target estimand, and generally falls outside the one-parameter family $\bm c(\omega)$, its excess asymptotic variance being given by \eqref{eq:gap} rather than \eqref{eq:excess}. The estimator remains consistent provided the fitted coefficient has a finite probability limit, because the between-arm covariate imbalance vanishes in probability; Table \ref{tab:natural} collects these estimators, their limiting coefficients, and the conditions under which each is optimal or less efficient than no adjustment.

\begin{table}[!ht]
    \centering
    \caption{The regression adjustments of \S\ref{sec:existing} and the proposed estimator as members of the class \eqref{eq:calibrated}. Each is $\widehat\lambda(a,1-a;\bm c)$ with the limiting coefficient shown, lying, where it lies on it, at position $\omega$ on the segment $\bm c(\omega)$ of Proposition \ref{prop:family}; the pairwise analysis of covariance lies off the segment.}
    \label{tab:natural}
    \setlength{\tabcolsep}{4pt}
    \small
    \begin{adjustbox}{max width=\linewidth}\begin{tabular}{llcll}
        \toprule
        Estimator & Coefficient $\bm c$ & $\omega$ & Optimal when & Worse than no adjustment when\\
        \midrule
        Pairwise, interacted \eqref{eq:pairlin} & $\tfrac12(\bm\gamma_1+\bm\gamma_2)$ & $\tfrac12$ & $\pi_a=\tfrac12$ or $\bm\gamma_1=\bm\gamma_2$ & $|\tfrac12-\pi_a|\,\|\bm\gamma_1-\bm\gamma_2\|>\|\bm c^{\ast}\|$\\
        Pairwise analysis of covariance \eqref{eq:pairancova} & involves $W_{i,j}(a',a')$ & none & not in general & determined by \eqref{eq:gap}\\
        Unit level, first role & $\bm\gamma_1$ & $0$ & $\bm\gamma_1=\bm\gamma_2$ & $\pi_a\|\bm\gamma_1-\bm\gamma_2\|>\|\bm c^{\ast}\|$\\
        Unit level, second role & $\bm\gamma_2$ & $1$ & $\bm\gamma_1=\bm\gamma_2$ & $\pi_{1-a}\|\bm\gamma_1-\bm\gamma_2\|>\|\bm c^{\ast}\|$\\
        Optimal calibration \eqref{eq:feasible} & $\pi_{1-a}\bm\gamma_1+\pi_a\bm\gamma_2$ & $\pi_a$ & always & never\\
        \bottomrule
    \end{tabular}\end{adjustbox}
    \par\vspace{4pt}\begin{minipage}{\linewidth}\footnotesize All norms are $\|\cdot\|_{\bm S_{XX}}$. The first four rows are the estimators of \S\ref{sec:existing}. The last row is the proposed estimator, whose coefficient is $\bm c^{\ast}$ in \eqref{eq:cstar}.\end{minipage}
\end{table}

\begin{remark}\label{sec:notinclass}
    We note that several approaches to covariate adjustment generally fall outside the linear calibration class in Lemma \ref{lem:class}, including probabilistic index regression and its extensions with a non-identity link function \citep{thas2012probabilistic,DeNeve2015,Mao2021}, nonlinear standardization \citep{Scheidegger2026}, and weighting or augmented weighting estimation \citep{Cao2025,Vermeulen2015,Vermeulen2023}. These approaches generally use fitted conditional models or pairwise weights rather than a linear correction for the between-arm covariate imbalance. The optimality result in Theorem \ref{thm:optimal} and the efficiency comparisons in Proposition \ref{prop:family} are hence restricted to the specified linear class and do not establish an efficiency ranking relative to these alternatives.
\end{remark}

\subsection{Constructing a feasible estimator for optimal covariate adjustment}\label{sec:feasible}

The coefficients \eqref{eq:gammas} involve position-specific scores that are only partly observed; $\overline\epsilon_{i,\cdot}(a,1-a)$ can be estimated only for units assigned to arm $a$ and $\overline\epsilon_{\cdot,i}(a,1-a)$ only for units assigned to arm $1-a$. To construct a feasible estimator, we define the observed scores
\begin{align*}
    \widehat W_{i,\cdot}=\frac{1}{N_{1-a}}\sum_{j:A_j=1-a}W_{i,j}\quad (A_i=a),\qquad
    \widehat W_{\cdot,i}=\frac{1}{N_{a}}\sum_{j:A_j=a}W_{j,i}\quad (A_i=1-a),
\end{align*}
and let $\widehat{\bm\gamma}_1(a,1-a)$ be the least squares coefficient of $\bm X_i$ in the regression of $\widehat W_{i,\cdot}$ on $(1,\bm X_i^\top)^\top$ over the units with $A_i=a$, and let $\widehat{\bm\gamma}_2(a,1-a)$ be minus the least squares coefficient of $\bm X_i$ in the regression of $\widehat W_{\cdot,i}$ on $(1,\bm X_i^\top)^\top$ over the units with $A_i=1-a$. Set $\widehat{\bm c}=\pi_{1-a}\widehat{\bm\gamma}_1(a,1-a)+\pi_a\widehat{\bm\gamma}_2(a,1-a)$ and the feasible estimator is
\begin{equation}\label{eq:feasible}
    \widehat\lambda^{\ast}(a,1-a)=\widehat\lambda(a,1-a)-\overline{\bm X}_{\mathcal S(a,1-a)}^\top\widehat{\bm c}.
\end{equation}

\begin{theorem}\label{thm:feasible}
    Suppose Conditions \ref{cd:design}--\ref{cd:nondeg} and the convergence hypothesis of Theorem \ref{thm:optimal} hold at $\bm c^\ast$, with $\sigma_\ast^2=\lim_{N\to\infty}V\{\overline r_i(a,1-a;\bm c^\ast)\}>0$. Then, with probability tending to one, both within-arm covariate second-moment matrices are nonsingular, so that $\widehat{\bm c}$ is well defined. Moreover, $\widehat{\bm c}-\bm c^\ast=O_{\mathbb P}(N^{-1/2})$ and $\widehat\lambda^\ast(a,1-a)-\widehat\lambda(a,1-a;\bm c^\ast)=O_{\mathbb P}(N^{-1})$. Consequently,
    \begin{align*}
        N^{1/2}\{\widehat\lambda^\ast(a,1-a)-\lambda(a,1-a)\}
        \ \xrightarrow{d}\ \mathcal N(0,\sigma_\ast^2).
    \end{align*}
    Thus, estimating the adjustment coefficient does not affect the first-order limiting distribution, and $\widehat\lambda^\ast(a,1-a)$ is asymptotically no less efficient than the unadjusted estimator.
\end{theorem}

Theorem \ref{thm:feasible} is proved in \S{S11}. Estimating $\bm c^{\ast}$ leaves the limiting distribution unchanged at first order. This is because $\widehat{\bm c}$ enters the estimator only through $\overline{\bm X}_{\mathcal S(a,1-a)}^\top\widehat{\bm c}$. Since $\overline{\bm X}_{\mathcal S(a,1-a)}=O_{\mathbb P}(N^{-1/2})$ under complete randomization, $\widehat\lambda^{\ast}(a,1-a)-\widehat\lambda(a,1-a;\bm c^{\ast})=-\overline{\bm X}_{\mathcal S(a,1-a)}^\top(\widehat{\bm c}-\bm c^{\ast})=O_{\mathbb P}(N^{-1})$ for any $\sqrt N$-consistent $\widehat{\bm c}$, and $\widehat\lambda^{\ast}(a,1-a)$ is therefore asymptotically no less efficient than the unadjusted estimator. Estimation does, however, replace the exact unbiasedness of Lemma \ref{lem:class}, which holds only for a nonrandom $\bm c$, with consistency. Since $\widehat\lambda(a,1-a)$ is unbiased, $\mathbb E(\overline{\bm X}_{\mathcal S(a,1-a)})=\bm 0$, and $\widehat{\bm c}$ and $\overline{\bm X}_{\mathcal S(a,1-a)}$ depend on the same assignment, we have
\begin{align*}
    \mathbb E\{\widehat\lambda^{\ast}(a,1-a)\}-\lambda(a,1-a)
    =-\sum_{k=1}^{d_x}{\rm cov}\big(\overline X_{\mathcal S(a,1-a),k},\ \widehat c_k-c_k^{\ast}\big),
\end{align*}
which need not vanish. Each term is a covariance between two $O_{\mathbb P}(N^{-1/2})$ quantities, so the bias is $O(N^{-1})$ provided $N\,\overline X_{\mathcal S(a,1-a),k}(\widehat c_k-c_k^{\ast})$ is uniformly integrable. This can fail even for a bounded contrast, because $\widehat{\bm c}$ involves the inverses of the within-arm covariate second-moment matrices, which, with a discrete covariate, are singular with positive probability at fixed $N$. On such an assignment, we take $\widehat{\bm c}$ to be the minimum-norm least squares solution, which defines \eqref{eq:feasible} everywhere and leaves Theorem \ref{thm:feasible} unaffected, that event having vanishing probability. A sufficient condition for the $O(N^{-1})$ bias is $\mathbb E\|\widehat{\bm c}-\bm c^{\ast}\|^2=O(N^{-1})$, by the Cauchy--Schwarz inequality. Theorem \ref{thm:feasible} does not provide this condition. Rather, it bounds $\widehat{\bm c}-\bm c^{\ast}$ in probability, which does not further bound a moment.

\section{Variance estimation and statistical inference}

\subsection{The proposed variance estimator}\label{sec:varest}

The first two terms of the leading randomization variance in \eqref{eq:leading} are within-arm second moments of residuals, and they can be estimated by within-arm sample moments. The coefficient $\widehat{\bm c}$ in \eqref{eq:feasible} combines the two arm-specific fits of \S\ref{sec:feasible}. For variance estimation, each arm uses its own fit. Let
\begin{align*}
    &\widehat r_{i}=\widehat W_{i,\cdot}-\widehat W_{(a)}-(\bm X_i-\overline{\bm X}_{(a)})^\top\widehat{\bm\gamma}_1\quad \text{if }A_i=a, \displaybreak[0]\\
    &\widehat s_{i}=\widehat W_{\cdot,i}-\widehat W_{(1-a)}+(\bm X_i-\overline{\bm X}_{(1-a)})^\top\widehat{\bm\gamma}_2\quad \text{if }A_i=1-a,
\end{align*}
be the residuals of those two fits, where $\widehat W_{(a)}=N_a^{-1}\sum_{i:A_i=a}\widehat W_{i,\cdot}$ and $\widehat W_{(1-a)}=N_{1-a}^{-1}\sum_{i:A_i=1-a}\widehat W_{\cdot,i}$ are the arm-specific means of the observed position-specific scores. The proposed variance estimator is
\begin{equation}\label{eq:varest}
    \widehat V=\frac{1}{\pi_a(N_a-d_x-1)}\sum_{i:A_i=a}\widehat r_i^{\,2}+\frac{1}{\pi_{1-a}(N_{1-a}-d_x-1)}\sum_{i:A_i=1-a}\widehat s_i^{\,2},
\end{equation}
each sum divided by the residual degrees of freedom of its fit rather than by the arm size, which leaves the limit unchanged and offsets the downward bias of least squares residuals in a small arm. For the difference contrast in Corollary \ref{cor:lin}, this is asymptotically equivalent to the heteroscedasticity-robust variance estimator of the fully interacted regression of \citet{Lin2013}, which is also built from the residuals of the two arm-specific regressions. It targets the finite-population counterpart, built from the residuals of the two regressions \eqref{eq:gammas},
\begin{equation}\label{eq:Vgamma}
    V_\gamma=\frac{1}{\pi_aN}\sum_{i=1}^N\overline e_{i,\cdot}^{\,2}+\frac{1}{\pi_{1-a}N}\sum_{i=1}^N\overline e_{\cdot,i}^{\,2},
\end{equation}
where $\overline e_{i,\cdot}=\overline\epsilon_{i,\cdot}(a,1-a)-\overline{\bm X}_{i,\cdot}^\top\bm\gamma_1$ and $\overline e_{\cdot,i}=\overline\epsilon_{\cdot,i}(a,1-a)+\overline{\bm X}_{i,\cdot}^\top\bm\gamma_2$.

\begin{theorem}\label{thm:varest}
    Under the conditions required in Theorem \ref{thm:feasible}, $\widehat V=V_\gamma+o_{\mathbb P}(1)$. Moreover,
    \begin{align*}
        V_\gamma-V\{\overline r_i(a,1-a;\bm c^{\ast})\}=\frac{1}{N}\sum_{i=1}^N\left(\overline e_{i,\cdot}+\overline e_{\cdot,i}\right)^2\geq 0,
    \end{align*}
    with equality if and only if the two residuals of every unit sum to zero. Consequently, $N^{-1}\widehat V$ is asymptotically conservative for the asymptotic variance $N^{-1}V\{\overline r_i(a,1-a;\bm c^{\ast})\}$ of $\widehat\lambda^{\ast}(a,1-a)$, and the confidence interval estimator $\widehat\lambda^{\ast}(a,1-a)\pm z_{1-\alpha/2}(\widehat V/N)^{1/2}$ has asymptotic coverage at least $1-\alpha$.
\end{theorem}

Theorem \ref{thm:varest} is proved in \S{S12}. The excess depends on the sum of a unit's two residuals. Only one of them is observed under a given assignment, and thus their joint variation is unidentifiable, as in \citet{Neyman1923} for the average treatment effect, where the excess is the finite-population variance of the individual treatment effects. One could instead form both residuals with the single coefficient $\widehat{\bm c}$ of \eqref{eq:feasible}. The next proposition shows that this alternative is more conservative than \eqref{eq:varest} by exactly $\|\bm\gamma_1-\bm\gamma_2\|^2_{\bm S_{XX}}$, the same difference that governs Proposition \ref{prop:family}.

\begin{proposition}\label{prop:bound}
    Let $\bm S_{XX}$ be nonsingular. The finite-population least squares coefficient of $\overline\epsilon_{i,\cdot}(a,1-a)+\overline\epsilon_{\cdot,i}(a,1-a)$ on $\overline{\bm X}_{i,\cdot}$ is $\bm\gamma_1-\bm\gamma_2$, and, with the arguments $(a,1-a;\bm c^{\ast})$ of the residuals suppressed,
    \begin{align}
        &V_c\{\overline r_i(a,1-a;\bm c^{\ast})\}-V_\gamma=\left\|\bm\gamma_1-\bm\gamma_2\right\|^2_{\bm S_{XX}}, \label{eq:bound-1}\displaybreak[0]\\
        &\frac{1}{N}\sum_{i=1}^N\left(\overline r_{i,\cdot}+\overline r_{\cdot,i}\right)^2=\frac{1}{N}\sum_{i=1}^N\left(\overline e_{i,\cdot}+\overline e_{\cdot,i}\right)^2+\left\|\bm\gamma_1-\bm\gamma_2\right\|^2_{\bm S_{XX}} \label{eq:bound-2}.
    \end{align}
    For any $\omega\neq\pi_a$, $\|\bm\gamma_1-\bm\gamma_2\|^2_{\bm S_{XX}}=\pi_a\pi_{1-a}(\omega-\pi_a)^{-2}[V\{\overline r_i(a,1-a;\bm c(\omega))\}-V\{\overline r_i(a,1-a;\bm c^{\ast})\}]$.
\end{proposition}

Proposition \ref{prop:bound} is proved in \S{S13}. The alternative estimator described before the proposition converges to $V_c\{\overline r_i(a,1-a;\bm c^{\ast})\}$, the first two terms in \eqref{eq:Vdef} at $\bm c^{\ast}$, as shown in \S{S15}. The coefficient $\bm c^{\ast}$ is a weighted average of $\bm\gamma_1$ and $\bm\gamma_2$. In arm $a$ the best-fitting coefficient is $\bm\gamma_1$, and in arm $1-a$ it is $\bm\gamma_2$. A single compromise coefficient therefore fits each arm worse than that arm's own coefficient does. By \eqref{eq:bound-1}, the total cost of the compromise is $\|\bm\gamma_1-\bm\gamma_2\|^2_{\bm S_{XX}}$. This cost is zero when the two projections agree and grows as they move apart. By the last statement of the proposition, the same quantity, multiplied by $(\omega-\pi_a)^2/(\pi_a\pi_{1-a})$, is the efficiency loss of the member $\bm c(\omega)$ of the family. The arm-specific residuals of \eqref{eq:varest} therefore help most in the populations where the unit-level adjustments lose the most efficiency.

\eqref{eq:bound-2} shows what is left. Both estimators are conservative because of the sum $\overline\epsilon_{i,\cdot}(a,1-a)+\overline\epsilon_{\cdot,i}(a,1-a)$ of a unit's two position-specific scores, which no assignment reveals. The covariates predict part of this sum, namely $\overline{\bm X}_{i,\cdot}^\top(\bm\gamma_1-\bm\gamma_2)$. The common-coefficient estimator is conservative by the second moment of the whole sum. The estimator \eqref{eq:varest} is conservative only by the second moment of the part the covariates cannot predict, $\overline e_{i,\cdot}+\overline e_{\cdot,i}$. This part is uncorrelated with $\overline{\bm X}_{i,\cdot}$, so no linear function of the covariates reduces it further.

The target $V_\gamma$ can still be tightened. The second-moment refinement of \citet{Aronow2014} applies to it, and \S{S16.4} evaluates the refinement exactly. In the design of \S\ref{sec:sim}, it lowers the conservatism factor of \eqref{eq:varest} by at most $0.009$.

\subsection{Estimation and inference on the summary estimands}\label{sec:netbenefit}

For the net benefit estimand $\tau(a)=\lambda(a,1-a)-\lambda(1-a,a)$, reversing the order of each pair gives $\overline{\bm X}_{\mathcal S(1-a,a)}=-\overline{\bm X}_{\mathcal S(a,1-a)}$, so that calibrating the two components with coefficients $\bm c$ and $\widetilde{\bm c}$ gives $\widehat\tau(a;\bm c,\widetilde{\bm c})=\widehat\tau(a)-\overline{\bm X}_{\mathcal S(a,1-a)}^\top(\bm c+\widetilde{\bm c})$. We define
\begin{align*}
    \overline\eta_{i,\cdot}=\overline\epsilon_{i,\cdot}(a,1-a)-\overline\epsilon_{\cdot,i}(1-a,a),\qquad
    \overline\eta_{\cdot,i}=\overline\epsilon_{\cdot,i}(a,1-a)-\overline\epsilon_{i,\cdot}(1-a,a)
\end{align*}
for the two net position-specific scores of unit $i$. Every statement about $\tau(a)$ concerns both orderings. The net scores are differences of position-specific scores from the two orderings, both covered by Condition \ref{cd:moments}, so they satisfy the Lindeberg requirement of Condition \ref{cd:nondeg} automatically. We record it as Condition \ref{cd:reversed} to name the hypothesis used in this subsection.

\begin{condition}\label{cd:reversed}
    $\max_{1\leq i\leq N}(\overline\eta_{i,\cdot}^2+\overline\eta_{\cdot,i}^2)/N\longrightarrow 0$.
\end{condition}

Under Conditions \ref{cd:design}--\ref{cd:reversed}, and with the hypotheses of Theorems \ref{thm:optimal}--\ref{thm:varest} imposed on the net scores, namely a leading variance bounded away from zero and convergent together with the two within-arm second moments composing it, arguments in \S\ref{sec:optimal} apply with $\overline\epsilon_{i,\cdot}(a,1-a)$ and $\overline\epsilon_{\cdot,i}(a,1-a)$ replaced by $\overline\eta_{i,\cdot}$ and $\overline\eta_{\cdot,i}$. That is, the optimal coefficient is $\pi_{1-a}$ times the regression coefficient of $\overline\eta_{i,\cdot}$ on $\overline{\bm X}_{i,\cdot}$ plus $\pi_a$ times that of $\overline\eta_{\cdot,i}$ on $\overline{\bm X}_{\cdot,i}$, the guaranteed efficiency gain has the same form as in Theorem \ref{thm:optimal}, and \eqref{eq:varest} applies with $\widehat r_i$ and $\widehat s_i$ computed from the observed net position-specific scores, namely $N_{1-a}^{-1}\sum_{j:A_j=1-a}(W_{i,j}-W_{j,i})$ for a unit with $A_i=a$ and $N_{a}^{-1}\sum_{j:A_j=a}(W_{j,i}-W_{i,j})$ for a unit with $A_i=1-a$. When the contrast is antisymmetric, $\overline\eta_{i,\cdot}=2\overline\epsilon_{i,\cdot}(a,1-a)$ and $\overline\eta_{\cdot,i}=2\overline\epsilon_{\cdot,i}(a,1-a)$, so that the optimal net benefit coefficient is $2\bm c^{\ast}$ and the asymptotic variance is four times that of $\widehat\lambda(a,1-a;\bm c^{\ast})$.

\begin{remark}\label{rmk:winratio}
    The win ratio $\lambda(a,1-a)/\lambda(1-a,a)$ can be handled by the same device. Write $\lambda_1=\lambda(a,1-a)$ and $\lambda_2=\lambda(1-a,a)$, both assumed bounded away from zero, and calibrate the two components with $\bm c$ and $\widetilde{\bm c}$. Since $\overline{\bm X}_{\mathcal S(1-a,a)}=-\overline{\bm X}_{\mathcal S(a,1-a)}$, a first-order expansion of the logarithm gives
    \begin{align*}
        \log\frac{\widehat\lambda_1(\bm c)}{\widehat\lambda_2(\widetilde{\bm c})}-\log\frac{\lambda_1}{\lambda_2}
        =\frac{\widehat\lambda_1-\lambda_1}{\lambda_1}-\frac{\widehat\lambda_2-\lambda_2}{\lambda_2}
        -\overline{\bm X}_{\mathcal S(a,1-a)}^\top\Big(\frac{\bm c}{\lambda_1}+\frac{\widetilde{\bm c}}{\lambda_2}\Big)+O_{\mathbb P}(N^{-1}),
    \end{align*}
    which becomes a family indexed by the single coefficient $\bm d=\lambda_1^{-1}\bm c+\lambda_2^{-1}\widetilde{\bm c}$. Provided that $\liminf_{N\to\infty}\min(\lambda_1,\lambda_2)>0$, and that the leading variance of the weighted net scores below is bounded away from zero and convergent, together with the two within-arm second moments composing it, \S\ref{sec:optimal} applies to the logarithm of the win ratio with the net position-specific scores of this subsection replaced by their inverse-$\lambda$ weighted counterparts $\lambda_1^{-1}\overline\epsilon_{i,\cdot}(a,1-a)-\lambda_2^{-1}\overline\epsilon_{\cdot,i}(1-a,a)$ and $\lambda_1^{-1}\overline\epsilon_{\cdot,i}(a,1-a)-\lambda_2^{-1}\overline\epsilon_{i,\cdot}(1-a,a)$. The optimal coefficient has the same form, the efficiency gain has the same form as in Theorem \ref{thm:optimal}, and \eqref{eq:varest} applies to the correspondingly weighted observed net position-specific scores. The derivation is in \S{S14}. The inverse weights $\lambda_1^{-1}$ and $\lambda_2^{-1}$ in $\bm d$ are unknown but can be estimated at a $\sqrt N$-rate by $\widehat\lambda_1^{-1}$ and $\widehat\lambda_2^{-1}$, which, by arguments in \S\ref{sec:feasible}, do not reduce efficiency asymptotically. When the contrast is antisymmetric, the win ratio becomes $\lambda_1/(C-\lambda_1)$, a monotone function of $\lambda_1$ alone, which, again, can be fully addressed via \S\ref{sec:optimal}; the win odds in \S\ref{sec:examples} is this case.
\end{remark}

A linear working model in the difference $D_{i,j}=A_i-A_j$ over all ordered pairs, the identity-link member of the family of \citet{Zhang2019}, estimates the net benefit through twice its slope and, with the covariate difference $\bm X_{i,j}$ added as a main effect, stays within the class of Lemma \ref{lem:class} with a coefficient of the analysis-of-covariance type; \S{S14} gives the details and the relation to the probabilistic index models of \citet{thas2012probabilistic}.

\begin{remark}\label{rmk:sandwich}
    Because the pair-level regressions \eqref{eq:pairancova} and \eqref{eq:pairlin} are least squares fits over the $N(N-1)$ ordered pairs, it is natural to estimate their variances by a sandwich formula, but the standard choices of the middle matrix are not in general valid here. Specifically, the heteroscedasticity-robust choice retains only the pairs identical in both indices and understates the variance by an order of magnitude, and clustering on one index omits the pairs in which a unit appears in the other position. The two-way cluster-robust estimator of \citet{Cameron2011} consistently estimates, for a single directional comparison, the target $V_c\{\overline r_i(a,1-a;\bm c)\}$ of the common-coefficient variance estimator at the fit's own coefficient $\bm c$, but for the net benefit it sets to zero the covariance between $\widehat\lambda(a,1-a)$ and $\widehat\lambda(1-a,a)$, which when $w$ is antisymmetric understates that target by a factor of two, so that its interval undercovers unless the finite population is conservative enough to offset the omission; adding the within-unit cross products restores the target. \S{S15} defines the four estimators for the fit on the two directional indicators with a fixed adjustment coefficient, proves these statements, and notes that for the fits \eqref{eq:pairancova} and \eqref{eq:pairlin} they hold to first order.
\end{remark}

\section{Numerical results and data application}

\subsection{Simulation study}\label{sec:sim}

We conducted a collection of simulation studies, with one design reported here and the remaining studies reported in \S{S16} of the Supplementary Material. In this design, a single parameter controls the discordance between the two position-specific projection coefficients, which sets the guaranteed gain in Theorem \ref{thm:optimal}, the excess variance of the family $\bm c(\omega)$ in Proposition \ref{prop:family}, and the gap between \eqref{eq:varest} and the common-coefficient variance estimator in Proposition \ref{prop:bound}. Thus, we check these results in one design by varying the parameter. Let $X_{1i} \sim \mathcal N(0,1)$, $X_{2i} \sim \mathcal B(1/2)$ independent of $X_{1i}$, $\zeta_i \sim \mathcal N(0,1)$ be the frailty shared by a unit's two potential outcomes, and $\varepsilon_{i1}$ and $\varepsilon_{i0}$ be independent centered exponential variables of scale $1/2$. The potential outcomes are generated from
\begin{align*}
    Y_i(1)=\frac{3}{10}+\frac{6}{5}X_{1i}+\frac{2}{5}X_{2i}+\zeta_i+\varepsilon_{i1},\qquad
    Y_i(0)=\theta\,\frac{6}{5}X_{1i}+\frac{2}{5}X_{2i}+\zeta_i+\varepsilon_{i0},
\end{align*}
so that the first covariate has slope $\tfrac65$ in the treated arm and $\theta\tfrac65$ in the control arm. The comparison function is $w(y_1,y_2)=\ind(y_1>y_2)$ and the estimand is $\lambda(1,0)$. We take $\theta\in\{1,1/2,0,-1/2,-1\}$, so that $\theta=1$ makes the covariates equally prognostic in the two arms and $\theta=-1$ makes them oppositely prognostic. We draw the finite population once for each sample size and then hold it fixed, and we draw $R=1{,}000$ assignment vectors by complete randomization with $N_1=\lceil\pi_1N\rceil$, using the same $R$ assignments for every estimator.

Table \ref{tab:theory} requires no simulation. Its entries are exact functionals of the realized finite population, obtained from \eqref{eq:quadratic}, \eqref{eq:gap} and \eqref{eq:excess}. At $\theta=1$ the two projection coefficients are close in the realized population, so the segment $\bm c(\omega)$ is very short, and every entry in that row agrees to the reported precision. At equal allocation, the fully interacted pairwise fit sits at $\omega=1/2=\pi_1$ and therefore attains the optimum at every $\theta$, while the two unit-level fits sit at the endpoints and are twice as variable as no adjustment when $\theta=-1$. Away from equal allocation, the pairwise fits lose their protection, such that at $\pi_1=0.7$ both are worse than no adjustment once $\theta\leq-1/2$. The row $\theta=0$ carries the practical message. A covariate that predicts the treated outcome and nothing else costs the unit-level fit a factor of $1.33$ at $\pi_1=0.7$, whereas the optimal coefficient instead gains a factor of $1/0.87$.

\begin{table}[!ht]
    \centering
    \caption{Design 1. Asymptotic variance of each adjustment relative to the unadjusted estimator, by the discordance parameter $\theta$ and the allocation. Values above one indicate that adjustment has reduced efficiency.}
    \label{tab:theory}
    \begin{adjustbox}{max width=\linewidth}\begin{tabular}{lccccccccccc}
        \toprule
        & \multicolumn{5}{c}{$\pi_1=0.5$} & & \multicolumn{5}{c}{$\pi_1=0.7$}\\
        \cmidrule(lr){2-6}\cmidrule(lr){8-12}
        $\theta$ & PI & PC & U1 & U2 & OPT & & PI & PC & U1 & U2 & OPT\\
        \midrule
        $1$ & 0.48 & 0.48 & 0.48 & 0.48 & 0.48 & & 0.49 & 0.49 & 0.49 & 0.49 & 0.49\\
        $1/2$ & 0.60 & 0.60 & 0.64 & 0.64 & 0.60 & & 0.67 & 0.68 & 0.75 & 0.68 & 0.67\\
        $0$ & 0.75 & 0.75 & 0.95 & 0.95 & 0.75 & & 0.91 & 0.98 & 1.33 & 0.96 & 0.87\\
        $-1/2$ & 0.90 & 0.90 & 1.45 & 1.45 & 0.90 & & 1.08 & 1.29 & 2.16 & 1.20 & 0.98\\
        $-1$ & 0.98 & 0.98 & 2.05 & 2.05 & 0.98 & & 1.01 & 1.44 & 2.69 & 1.20 & 0.86\\
        \bottomrule
    \end{tabular}\end{adjustbox}
    \par\vspace{4pt}\begin{minipage}{\linewidth}\footnotesize PI, fully interacted pairwise regression \eqref{eq:pairlin}; PC, pairwise analysis of covariance \eqref{eq:pairancova}; U1 and U2, the unit-level regressions with $\omega=0$ and $\omega=1$; OPT, the optimally calibrated estimator of \eqref{eq:cstar}. Entries are $V\{\overline r_i(1,0;\bm c)\}/V\{\overline\epsilon_i(1,0)\}$ for the coefficient $\bm c$ of each column, evaluated on the realized finite population of size $N=500$.\end{minipage}
\end{table}

\begin{table}[!ht]
    \centering
    \caption{Design 1 at $N=500$. Empirical variance of each adjustment relative to the unadjusted estimator; for the feasible calibrated estimator, the ratio of the average estimated standard error from \eqref{eq:varest} to the empirical standard error and the coverage of the nominal $95\%$ interval; and the exact conservatism factor of \eqref{eq:varest} and of the estimator that uses the common coefficient $\widehat{\bm c}$ in both arms. Based on $R=1{,}000$ randomizations.}
    \label{tab:sim}
    \setlength{\tabcolsep}{4pt}
    \small
    \begin{adjustbox}{max width=\linewidth}\begin{tabular}{lccccccccc}
        \toprule
        & \multicolumn{5}{c}{Relative empirical variance} & \multicolumn{2}{c}{Calibrated interval} & \multicolumn{2}{c}{Conservatism factor}\\
        \cmidrule(lr){2-6}\cmidrule(lr){7-8}\cmidrule(lr){9-10}
        $\theta$ & PI & PC & U1 & U2 & CA & ASE/ESE & Coverage & Proposed & Common $\bm c$\\
        \midrule
        \multicolumn{10}{l}{\ \emph{Allocation $\pi_1=0.5$}}\\
        $1$ & 0.48 & 0.48 & 0.49 & 0.48 & 0.48 & 1.10 & 0.966 & 1.06 & 1.06\\
        $1/2$ & 0.60 & 0.60 & 0.63 & 0.66 & 0.60 & 1.04 & 0.957 & 1.06 & 1.09\\
        $0$ & 0.74 & 0.74 & 0.94 & 0.96 & 0.74 & 1.10 & 0.968 & 1.07 & 1.19\\
        $-1/2$ & 0.92 & 0.92 & 1.54 & 1.49 & 0.92 & 1.13 & 0.968 & 1.07 & 1.33\\
        $-1$ & 1.00 & 1.00 & 2.04 & 2.23 & 0.99 & 1.10 & 0.967 & 1.07 & 1.50\\[2pt]
        \multicolumn{10}{l}{\ \emph{Allocation $\pi_1=0.7$}}\\
        $1$ & 0.54 & 0.54 & 0.54 & 0.55 & 0.54 & 1.02 & 0.953 & 1.05 & 1.05\\
        $1/2$ & 0.67 & 0.68 & 0.75 & 0.70 & 0.67 & 1.09 & 0.963 & 1.05 & 1.08\\
        $0$ & 0.95 & 1.02 & 1.38 & 0.97 & 0.90 & 1.06 & 0.954 & 1.06 & 1.16\\
        $-1/2$ & 1.08 & 1.27 & 2.11 & 1.24 & 1.00 & 1.04 & 0.954 & 1.06 & 1.28\\
        $-1$ & 1.03 & 1.42 & 2.58 & 1.24 & 0.89 & 1.01 & 0.955 & 1.06 & 1.43\\
        \bottomrule
    \end{tabular}\end{adjustbox}
    \par\vspace{4pt}\begin{minipage}{\linewidth}\footnotesize PI, PC, U1 and U2 as in Table \ref{tab:theory}; CA, the feasible calibrated estimator \eqref{eq:feasible}. ASE, average estimated standard error; ESE, empirical standard error. The conservatism factors are $[V_\gamma/V\{\overline r_i(1,0;\bm c^{\ast})\}]^{1/2}$ and $[V_c\{\overline r_i(1,0;\bm c^{\ast})\}/V\{\overline r_i(1,0;\bm c^{\ast})\}]^{1/2}$, evaluated on the realized finite population; by Proposition \ref{prop:bound}, the squares of the two factors differ by $\|\bm\gamma_1-\bm\gamma_2\|^2_{\bm S_{XX}}/V\{\overline r_i(1,0;\bm c^{\ast})\}$.\end{minipage}
\end{table}

Table \ref{tab:sim} gives the empirical counterpart. Several observations emerge from the results. First, the feasible calibrated estimator behaves like the oracle. Specifically, its relative variance never exceeds one, and it matches the OPT column of Table \ref{tab:theory} to within $0.06$ at every configuration, so estimating $\bm c^{\ast}$ costs nothing detectable at this sample size. The four least-squares adjustments do not share that protection, and at $\pi_1=0.7$ with $\theta=-1$ the first unit-level fit is more than two and a half times as variable as no adjustment. Second, the conservatism of \eqref{eq:varest} is as described in Theorem \ref{thm:varest}. The ratio of the average estimated standard error to the empirical one tracks the exact factor $[V_\gamma/V\{\overline r_i(1,0;\bm c^{\ast})\}]^{1/2}$ to within $0.06$ at every configuration, and that factor stays between $1.05$ and $1.07$ over the whole range of $\theta$. Coverage lies between $0.953$ and $0.968$, so the interval is valid at every configuration and only mildly cautious. Third, Proposition \ref{prop:bound} explains the gap between \eqref{eq:varest} and the estimator that uses the common coefficient $\widehat{\bm c}$ in both arms. At $\theta=1$, where the two coefficients nearly coincide, the two estimators have the same conservatism factor. At $\theta=-1$ and $\pi_1=0.5$, the factor of the common-coefficient estimator rises to $1.50$. The difference between the two targets is exactly $\|\bm\gamma_1-\bm\gamma_2\|^2_{\bm S_{XX}}$.

Six further simulation designs are considered in \S{S16}. Design 2 is the concordant case $\theta=1$ of the present design, in which $\bm\gamma_1$ and $\bm\gamma_2$ nearly coincide, so the segment $\bm c(\omega)$ is short, and all adjusted estimators are expected to perform alike. Design 3 uses composite and prioritized outcomes with $Q=2$, one component being ordinal with ties, under three comparison functions, to confirm that the results do not depend on the dimension or the type of the outcome. Design 4 uses the margin rule $w(y_1,y_2)=\ind(y_1>y_2+\delta)$, which is not antisymmetric, so that $\tau(1)$ carries information beyond $\lambda(1,0)$ and the net benefit estimators of \S\ref{sec:netbenefit} are tested on their own. Design 5 enlarges the working basis with quadratic and interaction terms, to measure the gain from a richer basis and to confirm that the gain is nondecreasing in its span. Design 6 compares the variance estimator \eqref{eq:varest} with the four pair-level sandwich estimators of Remark \ref{rmk:sandwich}, for $\lambda(1,0)$ and for $\tau(1)$. Design 7 makes the covariates independent of both potential outcomes and varies their number, so that no gain is available asymptotically and the finite-sample cost of estimating $\bm c^{\ast}$ is isolated. That section also reports every configuration of the present design at $N\in\{50,100,200,500\}$ and $\pi_1\in\{0.5,0.6,0.7,0.8\}$, and a study of the sensitivity of these conclusions to the draw of the finite population.

\subsection{Illustrative data application}\label{sec:app}

We illustrate the estimators on the Best Apnea Interventions for Research trial, an individually randomized study of continuous positive airway pressure in patients with obstructive sleep apnea \citep{Zhao2017}. We use the $120$ participants seen at the month $6$ visit with both outcomes and all covariates recorded, of whom $63$ received continuous positive airway pressure and $57$ received conservative medical therapy. The two outcomes are weighted twenty-four-hour systolic blood pressure and the Epworth Sleepiness Scale total, a win being the lower value of either. The covariates are age, sex, race, and trial site. We report four comparison functions: the Heaviside score on blood pressure alone; a margin rule on blood pressure requiring a lead of $5$ mmHg, which is not antisymmetric; the equally weighted rule on both outcomes; and the lexicographic rule that prioritizes sleepiness.

\begin{table}[!ht]
    \centering
    \caption{Best Apnea Interventions for Research trial, month $6$, $N=120$ with $\pi_1=0.525$. Estimates of $\lambda(1,0)$ and of the net benefit $\tau(1)$ under four comparison functions, unadjusted and optimally calibrated, with standard errors from \eqref{eq:varest}, applied without covariates for the unadjusted estimator.}
    \label{tab:app}
    \begin{adjustbox}{max width=\linewidth}\begin{tabular}{lccccccccc}
        \toprule
        & \multicolumn{4}{c}{$\lambda(1,0)$} & & \multicolumn{4}{c}{$\tau(1)$}\\
        \cmidrule(lr){2-5}\cmidrule(lr){7-10}
        & \multicolumn{2}{c}{Unadjusted} & \multicolumn{2}{c}{Calibrated} & & \multicolumn{2}{c}{Unadjusted} & \multicolumn{2}{c}{Calibrated}\\
        \cmidrule(lr){2-3}\cmidrule(lr){4-5}\cmidrule(lr){7-8}\cmidrule(lr){9-10}
        Comparison function & Est. & SE & Est. & SE & & Est. & SE & Est. & SE\\
        \midrule
        Blood pressure & 0.601 & 0.0526 & 0.606 & 0.0532 & & 0.202 & 0.1051 & 0.212 & 0.1065\\
        With margin & 0.481 & 0.0540 & 0.487 & 0.0554 & & 0.195 & 0.0997 & 0.205 & 0.1014\\
        Both outcomes & 0.590 & 0.0392 & 0.595 & 0.0387 & & 0.180 & 0.0784 & 0.189 & 0.0775\\
        Lexicographic & 0.587 & 0.0528 & 0.591 & 0.0510 & & 0.173 & 0.1056 & 0.183 & 0.1020\\
        \bottomrule
    \end{tabular}\end{adjustbox}
    \par\vspace{4pt}\begin{minipage}{\linewidth}\footnotesize Est., estimate; SE, standard error. Calibrated, the feasible calibrated estimator \eqref{eq:feasible} for $\lambda(1,0)$ and its net benefit counterpart of \S\ref{sec:netbenefit} for $\tau(1)$. The margin rule is not antisymmetric; under the other three, $\widehat\tau(1)=2\widehat\lambda(1,0)-1$.\end{minipage}
\end{table}

A treated participant has the lower blood pressure in an estimated $0.601$ of comparisons, and calibration moves this to $0.606$ with a standard error $1.3\%$ larger. Across the four comparison functions, calibration changes the standard error by between $-3.4\%$, under the lexicographic rule, and $+2.5\%$, under the margin rule. Changes of this size are what demographic covariates that explain little of the position-specific scores should deliver. Specifically, Theorem \ref{thm:optimal} guarantees only that the asymptotic gain is nonnegative, and at $N_1=63$, $N_0=57$ and $d_x=5$ the degrees-of-freedom correction in \eqref{eq:varest} alone inflates the estimated standard error by between $4.3\%$ and $4.8\%$ relative to the divisors $N_1-1$ and $N_0-1$ used without covariates. The point estimates nonetheless move. For blood pressure, the unadjusted interval for the net benefit is $(-0.004,0.409)$, and the calibrated interval is $(0.003,0.421)$, so the two differ in whether they exclude zero; the same happens under the margin rule, where the intervals are $(-0.001,0.390)$ and $(0.006,0.403)$. Under the lexicographic rule, both intervals contain zero, and under the equally weighted rule, neither does.

The quantities of \S\ref{sec:existing} can be evaluated on these data. For all four comparison functions, $\widehat{\bm\gamma}_1^\top\bm S_{XX}\widehat{\bm\gamma}_2>0$, so the covariates move a unit's two position-specific scores in opposite directions. Evaluating \eqref{eq:hurts} at the realized $\pi_1=0.525$ then shows that no member of the family $\bm c(\omega)$ is less efficient than no adjustment on this population. The unit-level fit is nonetheless wasteful, as it gives away between $0.17$ and $0.54$ of the gain the optimal coefficient achieves, and at $\pi_1=0.8$ it would give away between $0.33$ and $0.79$.

\S{S17} reports the full results under all four comparison functions, the estimated coefficients $\widehat{\bm\gamma}_1$ and $\widehat{\bm\gamma}_2$ covariate by covariate, and the criterion \eqref{eq:hurts} evaluated at three allocations.

\section{Discussion}\label{sec:discussion}

Growing interest in generalized pairwise comparisons for ordinal, multivariate, and prioritized outcomes has sharpened the question of how covariates should be used to improve precision \citep{Jaspers2024,Buyse2025,Scheidegger2026,Cao2025}. The primary contribution of this work is to provide a finite-population causal foundation for answering that question under complete randomization, with potential outcomes and covariates fixed and treatment assignment the sole source of randomness. The results connect to the design-based literature exemplified by \citet{Lin2013}, whose fully interacted least squares adjustment for the average treatment effect is asymptotically no less efficient than the unadjusted difference in means without requiring a correctly specified outcome model. We show that this guarantee extends to nonlinear pairwise estimands, but does not follow automatically from analogous regression specifications; that is, seemingly natural pair-level and single-arm adjustments can be less efficient than no adjustment. We therefore propose a novel calibration estimator that accounts for the treatment allocation and minimizes asymptotic randomization variance within the linear adjustment class, thereby restoring the efficiency guarantee for generalized pairwise comparisons. 

Our results on the randomization-based variance help explain why some familiar regression adjustments can reduce precision and how calibration avoids this problem. For pairwise comparisons, each unit has two average scores, corresponding to its two possible positions in the comparison. The optimal adjustment uses covariates to predict a weighted combination of these scores, with the weights determined by the treatment allocation. The pair-level and single-arm regression adjustments considered here generally use different weights, which can even lead to a loss of efficiency relative to no adjustment. The proposed calibration estimator uses the allocation to choose the appropriate weights, and estimating its adjustment coefficient does not affect the first-order variance. For comparisons based on outcome differences, the resulting estimator turns out to be equivalent to the estimator of \citet{Lin2013} to first order. For more general comparisons, these results clarify what covariates need to predict to improve precision. Their value depends on the comparison rule and the treatment allocation, so covariates that predict the outcome well may not be the most useful for a particular pairwise estimand.

Finally, the proposed variance estimator supports asymptotically conservative inference. The squared norm of the difference between the covariate projection coefficients for these scores is also what a variance estimator built on a single common coefficient pays over the proposed one, connecting the limitations of variance estimation to the efficiency losses of the regression adjustments. Extending these results to censored outcomes is an important direction for future work, since censoring can prevent some pairwise comparisons from being determined. Approaches such as inverse probability of censoring weighting, or the conditional tie weighting of \citet{fang2026making}, would require additional identification assumptions and theory that account for estimating the censoring distribution. The current results provide a foundation for these extensions by showing how covariate adjustment can improve precision for nonlinear pairwise estimands while retaining guarantees based on randomization.

\section*{Acknowledgement}
This work is supported by the United States National Institutes of Health (NIH), National Heart, Lung, and Blood Institute (NHLBI, grant number 1R01HL178513). All statements in this report, including its findings and conclusions, are solely those of the authors and do not necessarily represent the views of the NIH. 

\section*{Supplementary material}
\label{SM}

The Supplementary Material contains the exact moment computations in \S{S2}; all proofs in \S\S{S3}--{S13}, including the derivation of Remark \ref{rmk:exact} in \S{S8}; the net benefit and the win ratio in \S{S14}; the comparison with pair-level sandwich estimators in \S{S15}; the specification of all seven simulation designs with the results for every configuration and a sensitivity study in \S{S16}; and the full analysis of the Best Apnea Interventions for Research trial in \S{S17}.

\singlespacing
\bibliographystyle{jasa3}
\bibliography{PIM}

\end{document}